\documentclass[preprint,aps,superscriptaddress,nofootinbib]{revtex4-1}
\usepackage{graphicx}
\usepackage{braket}
\newcommand{\del}{\partial}

\begin{document}
\title{Dark energy from the gravity vacuum}

\author{Sandipan Sengupta}
\email{sandipan@iucaa.ernet.in}
\affiliation{Inter-University Centre for Astronomy and Astrophysics (IUCAA), Pune-411007, INDIA}

\begin{abstract}
We propose a new solution to the cosmological constant problem building on a nonperturbative quantum theory of gravity with torsional instantons. These pseudoparticles, which were recently found to exist in a first order formulation of Giddings-Strominger axionic gravity, carry nontrivial Nieh-Yan topological charge. The nonperturbative ground state as generated due to tunneling effects is shown to be stable under quantum fluctuations. Within this framework, the associated vacuum angle, namely the Barbero-Immirzi topological parameter, gets fixed to a numerical value determined by the Hubble constant.
\end{abstract}
  
 \preprint{IUCAA-13/2015} 
\maketitle

\section{Introduction}

In four dimensional gravity theory, there is a freedom to include three topological densities in the Lagrangian without affecting the classical dynamics. These are known as the Euler, Pontryagin and Nieh-Yan densities \cite{hehl,nieh,date,kaul,perez}. Whereas the first two depend on the curvature tensor, the third depends only on torsion \cite{nieh}. As in gauge theories, these topological terms can be expected to have a bearing on the quantum theory of gravity through potential instantonic effects in the Euclidean formulation \cite{raja}. And indeed, such expectations have been found to be realized with the discovery of gravitational instantons carrying Pontryagin and Euler charges \cite{hawking2,eguchi1,gibbons,smilga}. However, there has been no known example of a Nieh-Yan instanton which solves the classical equations of motion of Euclidean gravity with or without matter, until recently \cite{kauls}. In this work, it was shown that the Giddings-Strominger wormhole configurations \cite{gidd,gidd1}, which are solutions of a second order theory of axionic gravity, can be interpreted as torsional pseudoparticles carrying nontrivial Nieh-Yan topological numbers in a first order formulation of the same theory. Remarkably, the vacuum angle $\eta$ associated with this nonperturbative quantum theory was identified as the famous Barbero-Immirzi parameter of Loop Quantum Gravity \cite{barbero,immirzi}. In the effective Lagrangian, this appears as the coupling constant multiplying the Nieh-Yan topological density \cite{date,kaul,sengupta1}. To emphasize, the work in \cite{kauls} provides, for the first time, a clear demonstration of the fact that there exists an exact similarity between gauge theories with the famous $\theta$-vacuum and gravity theory with the hitherto unexplored $\eta$-vacuum.

Here, to begin with, we analyse the role of (spherically symmetric) quantum fluctuations around the torsional instantons in the semi-classical path integral formulation. Such an exercise is necessary in order to understand the stability properties of the nonperturbative $\eta$-vacuum against the emission or absorption of baby universes. 
We find that the quadratic fluctuation operator does not have any negative eigenmode. Subsequently, its determinant is evaluated and the expression of the ground state energy in terms of the Barbero-Immirzi angle $\eta$ is obtained.

Next, as a solution to the cosmological constant problem, we propose a framework where the energy of the $\eta$-vacuum has a natural interpretation as the dark energy that is believed to be responsible for the accelerated expansion of the universe at present. In this formulation, the naive perturbative (degenerate) ground states with all possible Nieh-Yan numbers are seen as the zero energy states of gravity theory, whereas the $\eta$-vacuum state with a nonvanishing energy $F_{\eta}$ is interpreted as the true ground state of our universe. As a consequence of our proposal, the Barbero-Immirzi vacuum angle gets completely determined in terms of the Newton's constant and Hubble parameter.

We note that the idea of applying wormhole physics to solve the cosmological constant problem has appeared earlier in the work of Coleman \cite{coleman} and has been discussed by many others \cite{hawking1,kleb,wein,jnan}. However, the essence of our proposal 
is very different from Coleman's which predicts an exactly vanishing cosmological constant. His mechanism has been invoked using the peakedness properties of the wavefunction of the universe, which is assumed to have the topology of a four-sphere. There have also been earlier discussions on the possibility of obtaining a cosmological constant within the context of $SU(2)$ gauge theories \cite{yoko}. The underlying assumption there is that the universe is yet to settle in a nonperturbative ground state of SU(2) gauge theory, and the true vacuum corresponds to a CP symmetric state with a vanishing $\theta$ angle and vanishing energy. Again, such studies have significant qualitative differences with our framework. The cosmological constant problem has also been discussed from various other perspectives, details of which can be found in ref.\cite{wein,copeland,martin,paddy,naresh}.

In what follows next, we briefly review the essential details of the torsional instanton and the $\eta$-vacuum in quantum gravity \cite{kauls}, before going into the main content of this article. 
  
\section{$\eta$-vacuum in quantum gravity}
We begin our discussion by writing down the first order Lagrangian density for (Euclidean) axionic gravity, following \cite{kauls}:
\begin{eqnarray}\label{LH} 
L(e,\omega,B)=-\frac{1}{2\kappa^2}e e^{\mu}_{I}e^{\nu}_{J}R_{\mu\nu}^{~IJ}(\omega)+\frac{1}{2\kappa}eH^{\mu\nu\alpha}e_{\mu}^I D_{\nu}(\omega) e_{\alpha I}
+\beta e H^{\mu\nu\alpha}H_{\mu\nu\alpha}
\end{eqnarray} 
where $B_{\mu\nu}$ is the antisymmetric tensor gauge field of rank two and $R_{\mu\nu}^{~IJ}(\omega)=\del_{[\mu}\omega_{\nu]}^{~IJ}+\omega_{[\mu}^{~IL}
\omega_{\nu]L}^{~~~J}$, $D_{\mu}(\omega)e_{\nu}^I=\del_{\mu}e_{\nu}^I+\omega_{\mu}^{~IJ}e_{\nu J}$, $H_{\mu\nu\alpha}=\del_{[\mu}B_{\nu\alpha]}$.
%
Variation of (\ref{LH}) with respect to $\omega_{\mu}^{~IJ}$ shows that this first order theory has a nonvanishing torsion given in terms of the field strength:
\begin{eqnarray*}\label{torsion}
T_{\alpha\beta}^{~~I}\equiv \frac{1}{2}D_{[\alpha}(\omega)e_{\beta]}^{I}= -\frac{\kappa}{2}~e^{\mu I}H_{\alpha\beta\mu}
\end{eqnarray*}
Using the decomposition $\omega_{\mu}^{~IJ}=\omega_{\mu}^{~IJ}(e)+K_{\mu}^{~IJ}$ of spin-connection into a torsionless part $\omega_{\mu}^{~IJ}(e)$ and contortion $K_{\mu}^{~IJ}$ and then using the identity $K_{\mu\nu\alpha}=K_{\mu}^{~IJ}e_{\nu I}e_{\alpha J}=T_{\mu\alpha\nu}-T_{\nu\mu\alpha}-T_{\alpha\mu\nu}$, one can obtain an explicit expression for contortion:
\begin{eqnarray}\label{cont}
K_{\mu\nu\alpha}=\frac{\kappa}{2}~H_{\mu\nu\alpha}
\end{eqnarray}
As elucidated in \cite{kauls}, this theory admits the Giddings-Strominger wormholes as classical solutions. To see this, let us adopt the Giddings-Strominger ansatz:\begin{eqnarray}\label{metric}
&&ds^2=d\tau^2+a^2(\tau)[d\chi^2+\mathrm{sin}^2 \chi d\theta^2+\mathrm{sin}^2 \chi \mathrm{sin}^2 \theta d\phi^2]
\nonumber\\
&&H^{\tau ab}=0,~H^{abc}=\frac{1}{\sqrt g}\epsilon^{abc}h(\tau,\chi,\theta,\phi)
\end{eqnarray}
where $h(\tau,\chi,\theta,\phi)$ is a scalar function and $\epsilon^{\tau abc}=\epsilon^{abc}$ is a totally antisymmetric density on the three-sphere whose indices are lowered using the induced three-metric $g_{ab}$. The equation for $B_{\mu\nu}$ is solved for:\begin{eqnarray}\label{h}
h(\tau)=\frac{\kappa Q}{3!~a^3(\tau)}
\end{eqnarray}
where, $Q$ is the axion charge, defined as the integral of the field strength over a closed three-surface:
\begin{eqnarray}\label{charge-int}
\int d^3 x ~\epsilon^{abc}H_{abc}=2\pi^2 \kappa Q
\end{eqnarray}
The tetrad equation, on using the $B_{\mu\nu}$ and torsion equations, leads to an evolution equation for the scale factor $a(\tau)$ that describes the Giddings-Strominger wormhole configuration:
\begin{eqnarray}\label{wh}
\dot{a}^2(\tau) =1-\frac{\kappa^4 F_a^2 Q^2}{18 a^4(\tau)}
\end{eqnarray}
where, $F_a^2=\beta-\frac{1}{8}$.
This wormhole has a throat of radius $a_0=18^{-\frac{1}{4}}\kappa (F_a Q)^\frac{1}{2}$ and represents a tunneling between two $R^3$ geometries widely separated in time. On the other hand, the half-wormhole tunnels between two different spatial topologies $R^3$ and $R^3+S^3$, and is interpreted as an instanton.

For our analysis, the torsion equation (\ref{cont}) is of critical importance, as this is what allows the interpretation of the Giddings-Strominger (half)wormholes as torsional instantons in the first order theory \cite{kauls}. Each such pseudoparticle carries a nontrivial Nieh-Yan topological charge, which turns out to be exactly equal to the axion charge of the baby-universe created by the instanton:
\begin{eqnarray}\label{nieh-yan}
N_{NY}&=&-\frac{1}{\pi^2 \kappa^2}\int_{M^4} d^4 x~ \del_{\mu}\left[\epsilon^{\mu\nu\alpha\beta}e_{\nu}^I D_\alpha(\omega)e_{\beta}^I\right]\nonumber\\
&=&\frac{1}{2\pi^2 \kappa}\int_{S^3} d^3 x ~\epsilon^{abc}H_{abc}~=~ Q
\end{eqnarray}
In the second line above, the four dimensional integral has been reduced to an integral over the (compact) $S^3$ boundary. This could be done because in the Giddings-Strominger theory, the baby universes, which are topologically $S^3$, are the (only) compact boundaries of the four manifold $M^4$. Importantly, this charge is integer-valued, as can be understood by considering the homotopy mappings induced by the Nieh-Yan index.
Thus, there exists an infinite number of degenerate ground states in gravity theory, each of them characterised by a definite Nieh-Yan number $N_{NY}$. Also, each half-wormhole has a finite action:\begin{eqnarray}\label{action}
S ~=~\frac{\pi^3}{\sqrt{2}} F_a Q~=~3\pi^3\frac{a^2_0}{\kappa^2}
\end{eqnarray} 
This suggests that there is a nontrivial amplitude for tunneling between any two states of different Nieh-Yan numbers. This results in a nonperturbative $\eta$-vacuum built out of a superposition of all the perturbative ground states, where the Barbero-Immirzi parameter $\eta$ emerges as a vacuum angle:
\begin{eqnarray}
\ket{\eta}=\sum_{N_{NY}}e^{i\eta N_{NY}}\ket{N_{NY}} \label{qvac}
\end{eqnarray}
Treating the distribution of instantons and anti-instantons of unit charge to be sufficiently dilute in the four geometry, the
transtion amplitude in the $\eta$-vacuum can be written as:\begin{eqnarray}\label{etaamp}
\braket{\eta'| e^{-F T} |\eta} = A ~\delta(\eta - \eta') ~\exp\left[2e^{-S}KVT \mathrm{cos}\eta \right]
\end{eqnarray}
where $A$ is a normalisation factor and $K$ encodes the contribution from quantum fluctuations in the semi-classical path integral.
Importantly, this shows that there is a correction to the `vacuum energy density' due to tunneling:
\begin{eqnarray}\label{F}
\frac{F_{\eta}}{V}=-2e^{-S} K \mathrm{cos} \eta
\end{eqnarray}
However, no attempt was made in \cite{kauls} to compute $K$ explicitly. We take this up in the next section.

Note that there could be an additional vacuum angle in this theory, namely, the coefficient of the Pontryagin density in the effective gravity Lagrangian. This would be the case if the torsional instanton carries a nontrivial Pontryagin index given by:
\begin{eqnarray*}
N_P=\int d^4 x~ \epsilon^{\mu\nu\alpha\beta}R_{\mu\nu}^{~~IJ}R_{\alpha\beta IJ}~
\end{eqnarray*} 
However, for the `canonical' choice of the axion coupling given by $\beta=\frac{1}{6}$  or $F_a^2=\frac{1}{24}$ (defined through equations (\ref{LH}) and (\ref{wh}) respectively), the Pontryagin number is identically zero. From here on, we will assume this value of the axion coupling which implies that there is no P and T violating parameter other than the Barbero-Immirzi angle $\eta$ in the quantum theory as considered here. As a result, the wormhole size $a_0$ gets completely fixed in terms of the Planck length $\kappa$:
\begin{eqnarray}\label{size}
a_0= \frac{\kappa}{2.3^{\frac{3}{4}}}
\end{eqnarray}

\section{Quantum fluctuations}
The action (\ref{LH}), after using the equations of motion for $B_{\mu\nu}$ and contortion $K_{\mu\nu\alpha}$, reads:
\begin{eqnarray}\label{S}
S[a]=\frac{6\pi^2 a_{0}^2}{\kappa^2}\int d\tau~ \left[-\left(1+\dot{a}^2(\tau)\right)a(\tau)+\frac{1}{a^3(\tau)}\right]
\end{eqnarray}
In the above we have introduced dimensionless variables $\tau$ and $a(\tau)$ using a rescaling of the original variables:
\begin{eqnarray*}
\tau \rightarrow \frac{\tau}{a_0}, ~a(\tau)\rightarrow \frac{a(\tau)}{a_0}
\end{eqnarray*}
The second order variation of (\ref{S}) yields:
\begin{eqnarray}
\delta^2 S[a] =\frac{6\pi^2 a_{0}^2}{\kappa^2}\int d\tau~ \delta a(\tau) \left[a(\tau)\frac{d^2}{d \tau^2}+\dot{a}(\tau)\frac{d}{d \tau}
+\frac{8}{a^5(\tau)}\right] \delta a(\tau)
\end{eqnarray}
For convenience, we shall try to find the eigenvalues of the operator: 
\begin{eqnarray}\label{O}
\hat{O}= a(\tau)\left[a(\tau)\frac{d^2}{d \tau^2}+\dot{a}(\tau)\frac{d}{d \tau}+\frac{8}{a^5(\tau)}\right]~
\end{eqnarray} 
This is perfectly fine as long as a new integration measure $du$ is used for normalising the eigenmodes, with:
\begin{eqnarray}\label{measure}
d u=\frac{d\tau}{a(\tau)} ~
\end{eqnarray}
To emphasize, the denominator above is completely fixed by the choice of the prefactor at the right hand side of eq.(\ref{O}).
In terms of the dimensionless variables and the operator $\hat{O}$, the one-instanton contribution to the path integral in the semiclassical appproximation becomes:
\begin{eqnarray}\label{pi}
&&Ae^{-S}\int \frac{a_0}{\kappa}d[\delta a]~e^{-\frac{3\pi^2 a_0^2}{\kappa^2}\int d\tau~\delta a(\tau)\hat{O}\delta a(\tau)}=A e^{-S}KVT\nonumber\\
&&~
\end{eqnarray}
where the factor of spacetime volume $VT$ arises due to an integration over the instanton location in the four geometry.
 Next, let us redefine the scale factor $a(\tau)$ as:
\begin{eqnarray*}
\chi^2(\tau)=1-\frac{1}{a^4(\tau)}
\end{eqnarray*}
In terms of this variable, the eigenvalue equation for operator $\hat{O}$ becomes:
\begin{eqnarray}
4(1-\chi^2)\frac{d}{d \chi}\left[(1-\chi^2)\frac{d \psi}{d \chi}\right]+8(1-\chi^2)\psi=\lambda \psi
\end{eqnarray}
The only eigenmode which is finite at the boundaries is given by:
\begin{eqnarray*}
\psi(\chi)=\sqrt{2}~(1-\chi^2)^{\frac{1}{2}}
\end{eqnarray*}
This corresponds to a positive eigenvalue $\lambda=4$ and is normalized with respect to the appropriate measure defined in equation (\ref{measure}):
\begin{eqnarray*}
\int du ~|\psi|^2=1~.
\end{eqnarray*}
The other solutions, given by $\psi(\chi)=(1-\chi^2)^{-1},~\chi(1-\chi^2)^{-\frac{1}{2}},~\chi(1-\chi^2)^{-\frac{7}{2}}$ all diverge at the boundaries. 

Note that here we do not use the Gibbons-Hawking-Perry (GHP) rotation \cite{perry,perry1}, a prescription invoked sometimes to perform a gaussian integration when the quadratic fluctuation part has a `wrong' sign. This is contrary to an earlier analysis by Rubakov et al.\cite{rubakov} who perform a GHP rotation of the variables first and then find the eigenmodes. However, the fact that there is no compelling reason to do so in a theory other than pure gravity and that such a naive application of the GHP method might lead to misleading results have been emphasized earlier in several contexts \cite{mazur,polchinski}.

There is also a zero mode among the fluctuations due to the time translation invariance, and is given by $\psi_0=\dot{a}(\tau)$ upto a normalization. Since this eigenfunction does not vanish at the $R^3$ boundary, its normalization can be defined only after a regularization, namely, a subtraction of the divergent boundary contribution. In that case, the norm of the zero mode becomes equal to the dimensionless instanton action, exactly as in standard instanton physics \cite{raja}:
\begin{eqnarray}
\int_{reg} du ~|\psi_0|^2=\int du ~\left|\frac{da}{du}\right|^2 + \frac{1}{2}a^2 \dot{a}|_{R^3}=\frac{\pi}{4}
\end{eqnarray}
Thus, the integration over the zero mode is equivalent to an integration over the time location $\tau_0$ multiplied by the factor $\frac{\sqrt{\pi}}{2}$. As a result, the total contribution $K$ due to the quadratic fluctuations, as defined in (\ref{etaamp}) and (\ref{pi}), can be written as:
\begin{eqnarray}\label{K}
K=\left[\frac{\sqrt{\pi}}{2}.\frac{a_0}{\kappa}.\frac{\kappa}{\sqrt{3}\pi a_0}\right](Det~\hat{O})^{-\frac{1}{2}}=\frac{1}{4\sqrt{3\pi}}
\end{eqnarray}
Similarly, the integration over the three other zero modes corresponding to the spatial coordinates $\chi,\theta$ and $\phi$ is equivalent to an integration over the location of the instanton over the three volume. This is precisely the origin of the factor $V$ in the one-instanton path integral (\ref{pi}). Finally, equation (\ref{K}) allows us to obtain an exact expression of the vacuum energy density (\ref{F}) in terms of only one free parameter, i.e. the vacuum angle $\eta$:  
\begin{eqnarray}\label{lambda}
\braket{\eta|\left[\frac{F_{\eta}}{V}\right]|\eta}=-\frac{1}{2\sqrt{3\pi}}e^{-\frac{\pi^3}{4\sqrt{3}}} \mathrm{cos}\eta
\end{eqnarray}

\section{Solution of the cosmological constant problem} 
The instanton density $n$, which is same as the number of tunneling events per unit Planck volume, is of the order of $Ke^{-S}$. Using the explicit expressions for $K$ and $S$ as given in the previous section, this can be found to be:
\begin{eqnarray*}
n \sim 10^{-2} \kappa^{-4}
\end{eqnarray*} 
Thus, the average separation between the instantons is of the order of $10^2 \kappa^4$, which is much larger than the size of an instanton, which is roughly given by (from eq.(\ref{size})): 
\begin{eqnarray*}
a_0^4\sim 10^{-3}\kappa^4
\end{eqnarray*}
This clearly shows that the dilute gas approximation is reliable in this context. The corresponding action $S$ leads to an exponentially small vacuum energy density (\ref{lambda}) which is also independent of spacetime. From general arguments based on the principle of Lorentz invariance \cite{wein,martin}, such a constant energy density would always lead to an `effective' energy-momentum tensor of the form:
\begin{eqnarray*}
\braket{\eta|T_{\mu\nu}|\eta}=-\frac{F_{\eta}}{V} g_{\mu\nu}~,
\end{eqnarray*}
where it has been assumed that the expectation values above are evaluated in a Lorentz invariant manner. This implies that $F_{\eta}$ corresponds to a negative pressure. Remarkably, these are exactly the features which are expected to be shared by the dark energy, if we ignore the zero point fluctuations due to quantum fields. This motivates us to identify $F_{\eta}$ precisely as the dark energy or the effective cosmological constant. 

Notice that the presence of the factor `$cos\eta$' in (21) implies that the torsional instantons are capable of producing both a small and a large vacuum energy density, depending upon the value of the $\eta$-angle. However, here we would like to focus only on a scenario consistent with current observations, which favours the presence of a tiny amount of vacuum energy that is thought to be driving the accelerated expansion of the universe today. 

Let us assume that our universe has reached a stage where it is in the true ($\eta$) vacuum state of quantum gravity generated due to tunneling effects as opposed to being in a perturbative vacuum of zero energy. Then, (\ref{lambda}) must be equal to the energy density $\rho$ of the present universe. From observations, $\rho$ is known to be about 70 percent of the critical density $\rho_c=\frac{3H_0^2}{8\pi G}$, where $H_0$ is the Hubble constant. As an immediate consequence, the vacuum angle $\eta$ gets constrained as $\frac{\pi}{2} \leq \eta \leq \frac{3\pi}{2}$, since the energy density (\ref{lambda}) must be positive. The wormhole size $a_0$ which is fixed in terms of the Planck length as in (\ref{size}) should be understood as the cut-off scale of quantum gravity in this context.

Based on the interpretation of $F_{\eta}$ as the dark energy, we have the following equality:
 \begin{eqnarray}\label{lambda1}
&&-2e^{-S}K \mathrm{cos}\eta =\alpha e^{-276} \nonumber\\\mathrm{or,~}
&& ln(-\mathrm{cos}\eta)= S-276+ln\left[\frac{\alpha}{2 K}\right]
\end{eqnarray}
where, $\alpha\approx 0.7$. 
This leads to:
\begin{eqnarray}\label{lambda4}
&&ln\left(-\mathrm{cos}\eta\right)= -269.7 \nonumber\\\mathrm{or,~}
&&\eta = \frac{\pi}{2}+\mathrm{sin}^{-1}\left(e^{-269.7}\right)=\frac{\pi}{2}+\delta(H_0)
\end{eqnarray}
Thus, the Barbero-Immirzi parameter $\eta$ gets fixed to a value close to $\frac{\pi}{2}$ radians, provided the ground state energy receives no further contribution from any other vacuum angle in the quantum theory. The small deviation $\delta(H_0)$ is completely determined in terms of the current estimate of Hubble parameter (and Newton's constant).
Evidently, the above form of the constraint is hardly sensitive to the small quantum corrections $K$ around the instantons.
In a more general case where other vacuum angles might also contribute to the vacuum energy through potential instanton effects, the effective energy-momentum tensor should have the general form as below: 
\begin{eqnarray*}
\braket{T_{\mu\nu}}=-\sum_{i}\frac{F_{\theta_i}}{V}g_{\mu\nu}~,
\end{eqnarray*}
where $F_{\theta_i}$ is the vacuum energy depending on the `i'-th vacuum angle $\theta_i$.

\section{Concluding remarks}
Here we have proposed a solution to the cosmological constant problem   using the nonperturbative vacuum structure in quantum gravity. The vacuum (dark) energy in this formulation depends on the Barbero-Immirzi vacuum angle, which is a quantum coupling constant of gravity theory. In this analysis, we do not address the issue of the zero point energy of fields. However, even if these vacuum fluctuations are relevant, they can now be treated satisfactorily using counterterms (e.g. the cosmological constant $\Lambda$) in the Lagrangian. Such bare terms can be adjusted to cancel the zero point energy exactly and hence would not need any fine-tuning. To emphasize, we have demonstrated that the existence of the rich $\eta$-vacuum in quantum gravity is sufficient to explain the origin of the small but nonvanishing energy density driving the accelerated expansion of the present universe.


As an important consequence of our proposal, the dynamics of the universe naturally fixes the Barbero-Immirzi parameter to a numerical value determined by the Hubble constant\footnote{The only other context where it is possible to fix the value of Barbero-Immirzi parameter is loop gravity, namely, by using the Black hole entropy formula \cite{abhay,rom,romesh}.}. 
As mentioned already, it is also possible to have a large vacuum energy for a sufficiently large value of $|cos\eta|$. Whether or not it is possible to find a realization of such a scenario in the context of early universe physics is an issue that requires a deeper study.
It is crucial to note that the analysis here does not depend on the details of the source of torsion, i.e. on whether it is generated by axionic or other matter or by a purely geometric configuration (degenerate tetrad). 

Here our focus was on the effect of torsional instantons, which lead to a nonperturbative ($\eta$) vacuum structure in quantum gravity. In a more general scenario, additional parity violating vacuum angles (e.g. the coefficient of gravitational Pontryagin index) could feature in the quantum theory and the numerical constraint on $\eta$ is likely to change. Nevertheless, the framework as set up here would still be applicable, while leading to a more general constraint involving all such topological parameters. A complete analysis along these lines can reveal hitherto unnoticed but important relationships among the coupling constants of nature.  
   


\acknowledgments 
It is a pleasure to thank S Kar and R Kaul for numerous enlightening discussions. Critical comments of N Dadhich, T Padmanabhan and V Sahni on the manuscript are also gratefully acknowledged.

\end{document}